\documentclass[twocolumn,pra,twocolumn,numerical,superscriptaddress]{revtex4-2}
\usepackage[latin9]{inputenc}
\usepackage[a4paper]{geometry}
\usepackage{color}
\usepackage{float}
\usepackage{bm}
\usepackage{amsmath}
\usepackage{amssymb}
\usepackage{graphicx}
\usepackage[unicode=true,pdfusetitle,
 bookmarks=true,bookmarksnumbered=false,bookmarksopen=false,
 breaklinks=false,pdfborder={0 0 0},pdfborderstyle={},backref=false,colorlinks=true]
 {hyperref}
\hypersetup{
 linkcolor=blue, urlcolor=blue, citecolor=blue}

\makeatletter

\providecommand{\tabularnewline}{\\}

\usepackage{dcolumn}
\usepackage{bm}
\usepackage{epsfig}
\usepackage{braket}

\renewcommand{\fnum@figure}{FIG.~\thefigure}
\usepackage{upgreek}
\usepackage{placeins}
\usepackage{tikz}
\usepackage{pgfplots}
\pgfplotsset{compat=1.5}

\makeatother

\begin{document}
\title{Visualizing coherent molecular rotation in a gaseous medium}

\author{Ilia Tutunnikov}
\affiliation{AMOS and Department of Chemical and Biological Physics, The Weizmann Institute of Science, Rehovot 7610001, Israel}
\author{Emilien Prost}
\affiliation{Laboratoire Interdisciplinaire CARNOT de Bourgogne Franche-Comt\'e, UMR 6303 CNRS-Universit\'e de Bourgogne, BP 47870, 21078, Dijon, France}
\author{Uri Steinitz}
\affiliation{AMOS and Department of Chemical and Biological Physics, The Weizmann Institute of Science, Rehovot 7610001, Israel}
\affiliation{Soreq Nuclear Research Centre, Yavne 8180000, Israel}
\author{Pierre B\'ejot}
\affiliation{Laboratoire Interdisciplinaire CARNOT de Bourgogne Franche-Comt\'e, UMR 6303 CNRS-Universit\'e de Bourgogne, BP 47870, 21078, Dijon, France}
\author{Edouard Hertz}
\affiliation{Laboratoire Interdisciplinaire CARNOT de Bourgogne Franche-Comt\'e, UMR 6303 CNRS-Universit\'e de Bourgogne, BP 47870, 21078, Dijon, France}
\author{Franck Billard}
\affiliation{Laboratoire Interdisciplinaire CARNOT de Bourgogne Franche-Comt\'e, UMR 6303 CNRS-Universit\'e de Bourgogne, BP 47870, 21078, Dijon, France}
\author{Olivier Faucher}
\email[Corresponding author: ]{olivier.faucher@u-bourgogne.fr}
\affiliation{Laboratoire Interdisciplinaire CARNOT de Bourgogne  Franche-Comt\'e, UMR 6303 CNRS-Universit\'e de Bourgogne, BP 47870, 21078, Dijon, France}
\author{Ilya Sh. Averbukh}
\email[Corresponding author: ]{ilya.averbukh@weizmann.ac.il}
\affiliation{AMOS and Department of Chemical and Biological Physics, The Weizmann Institute of Science, Rehovot 7610001, Israel}

\begin{abstract}
Inducing and controlling the ultrafast molecular rotational dynamics using shaped laser fields is essential in numerous applications. Several approaches exist that allow following the coherent molecular motion in real-time, including Coulomb explosion-based techniques and recovering molecular orientation from the angular distribution of high harmonics.
We theoretically consider a non-intrusive optical scheme for visualizing the rotational dynamics in an anisotropic molecular gas. The proposed method allows determining the instantaneous orientation of the principal optical axes of the gas.
The method is based on probing the sample using ultra-short circularly polarized laser pulses and recording the transmission image through a vortex wave plate. We consider two example excitations: molecular alignment induced by an intense linearly polarized laser pulse and unidirectional molecular rotation induced by a polarization-shaped pulse. The proposed optical method is promising for visualizing the dynamics of complex symmetric- and asymmetric-top molecules.
\end{abstract}
\maketitle

\section{Introduction}

Over the years, the scientific field of molecular alignment and orientation control
generated an extensive set of tools, ranging from adiabatic/impulsive
alignment by single linearly polarized laser pulses to orientation
by combined laser fields with intricately tailored polarizations.
Molecular alignment and unidirectional rotation (UDR) have been reviewed
\citep{Stapelfeldt2003,Fleischer2012}. For broader reviews on molecular
control using electromagnetic fields, the reader is referred to \citep{Lemeshko2013,Koch2019}.
The problem of control comes hand in hand with the problem of visualization
of the resulting dynamics.

In the early days, mainly Coulomb explosion technique was utilized for visualizing laser-induced one dimensional molecular alignment. An intense time-delayed
probe pulse ionized the molecules, and the yield of fragments ejecting
along/against the polarization of the probe was detected  \citep{Normand1992,Dietrich1993}.
This allowed recovering the degree of alignment, usually quantified
by a single observable $\braket{\cos^{2}\theta}$, where the angle
brackets denote the average value. The angle $\theta$ is the
angle between the molecular axis and the polarization direction of
the aligning pulse. Later on, the approach evolved into a nowadays standard
technique---velocity map imaging (VMI), allowing reconstruction of
molecular angular distribution as a function of time \citep{Eppink1997}.
For a recent example of the state of the art experiment using VMI, the reader is referred to \citep{Karamatskos2019} and the
references therein.
Another powerful tool providing access to 3D information is
the cold target recoil ion momentum spectroscopy (COLTRIMS) \citep{Dorner2000},
where the electron and ion momenta are coincidentally detected.
Methods based on imaging of charged fragments were successfully applied for imaging the dynamics
of unidirectionally rotating molecules and orientation dynamics of
asymmetric top molecules \citep{Mizuse2015,Lin2015,Lin2018}.
While VMI and COLTRIMS provide a complete characterization of molecular rotation, this usually comes at a price of strict experimental conditions (rarefied gases/molecular beam), complex experimental setups, and long acquisition times.

Another class of methods relies on optical techniques which have many
attractive practical advantages, e.g., a much-extended working range
of pressures and temperatures \citep{Faucher2011}. So far, the optical
detection schemes have been limited to one-dimensional measurements
of the ensemble-averaged quantities, such as the degree of molecular
orientation or alignment, quantified by $\braket{\cos\theta}$ \citep{Fleischer2011,Damari2016}
and $\braket{\cos^{2}\theta}$ \citep{Renard2003,Damari2016}, respectively.
Higher-order moments $\braket{\cos^{n}\theta}$, with $n>2$, of the
molecular angular distribution can be measure using harmonic
generation \citep{Weber2013,Karras2016}. Recently, angle-resolved
high-order-harmonic spectroscopy was used for generating molecular
``rotational movies'' with the help of machine learning tools \citep{He2019}.

Here, we present a theoretical analysis of the recently demonstrated
\citep{Bert2019} purely optical approach allowing detecting the instantaneous
orientation of the principal optical axes of laser-excited molecular
gas. The approach relies on ultrafast birefringence measurement using
delayed femtosecond probe pulses. The paper is organized as follows:
Sec. \ref{sec:Qualitative-Homo} qualitatively describes the proposed
imaging technique in the case of homogeneously excited molecular sample.
In Section \ref{sec:Beam-Prop-Homo}, we present the numerical results
obtained by solving the paraxial beam propagation equation. We use
two example excitations: impulsive excitation by linearly polarized
pulses resulting in a molecular dynamics that toggle between alignment and antialignment, and excitation by femtosecond polarization-twisted pulses which cause unidirectional rotation of the alignment axis.
Sections \ref{sec:Qualitative-Inhomo} and \ref{sec:Beam-Propg-Inhomo}
describe the case of the inhomogeneous excitation and present the corresponding
numerical results, respectively. Section \ref{sec:Conclusions} concludes
the paper.

\section{Qualitative Description -- Planar wave case \label{sec:Qualitative-Homo}}

We begin with a qualitative description of the proposed optical imaging
approach. As an example, we consider the case of rigid linear molecules
in the gas phase excited by a short \emph{linearly polarized} in the
$XY$ plane pulse (pump pulse), propagating along the $Z$-axis. Such
an excitation results in a transient molecular alignment along the
polarization axis of the pump pulse \citep{Stapelfeldt2003,Fleischer2012,Lemeshko2013,Koch2019}.
Due to quantum revivals \citep{Averbukh1989,Robinett2004}, the alignment
recurs periodically with a well-defined period. Each revival event
consists of alignment and antialignment stages. During both stages,
the molecular gas develops anisotropy---the refractive index for
light polarized along the pump polarization axis differs from the
refractive index for light polarized along the orthogonal direction. 
The two orthogonal optical principal axes in the $XY$ plane, 
$X'$ and $Y'$, have refractive indices $n_{X'}$ and $n_{Y'}$,
respectively. In the case of linearly polarized pump pulse, we let
$X'$ to be the axis of the pump polarization. Here, the direction
of $X'$ and $Y'$ axes are fixed, while the values of $n_{X'}$ and
$n_{Y'}$ depend on the probe delay. We use a circularly polarized probe
pulse, whose polarization can be decomposed onto the $X'$ and $Y'$ axes. 

Since, generally $n_{X'}\neq n_{Y'}$, the two components of the probe 
light accumulate a relative phase as they propagate through the 
anisotropic medium. After passing the medium, the probe polarization becomes, therefore, elliptical. The orientation of the major axis of the ellipse depends
on the orientation of the optical principal axes. For a homogeneous
medium, the major axis of the polarization ellipse is at $\pm\pi/4$
to the optical axes. The sign depends on the sense of circular polarization
and the stage of the molecular dynamics (alignment/antialignment).
One way to extract the information encoded in the polarization of
the probe pulse is to use polarization axis finder (PAF) \citep{Moh2007,Lei2018},
which renders a spatial intensity pattern having the same directionality
as the polarization ellipse. In the present work, the PAF is composed of a vortex
plate acting as a radial polarizer \citep{Yamaguchi1989,Stalder1996,Kozawa2005,Erdelyi2008,Moreno2012} and
a linear polarizer. Recently, a similar PAF was successfully
applied to imaging of coherent molecular rotors \citep{Bert2019}.

\begin{figure}[h]
\begin{centering}
\includegraphics{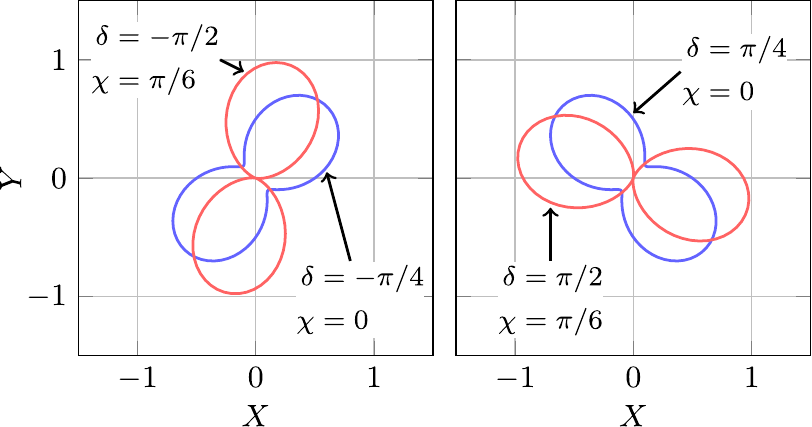}
\par\end{centering}
\caption{Polar plots of the intensity in Eq. \eqref{eq:intensity-afo-phi}. Left - alignment stage, $\delta<0$. Right - antialignment stage, $\delta>0$.
\label{fig:Homo-polar-plots}}
\end{figure}

Qualitatively, given a circularly polarized light at the input plane,
the electric field at an output plane can be determined with the help
of Jones calculus \citep{Hecht2016}. Here, we adopt the same phase
convention as in \citep{Hecht2016}, namely the phase of a monochromatic
plane wave is defined as $kz-\omega t$, where $k$ is the wave number,
$z$ is the position along the $Z$ axis, $\omega$ is the angular
frequency of light, and $t$ is time. At the input plane, the Jones
vector (expressed in the $X'Y'$ basis) of left-circular (from the
point of view of the receiver) probe light propagating along the $Z$
axis reads
\begin{equation}
\mathbf{u}_{X'Y'}(z_{i})=\frac{1}{\sqrt{2}}\left(\begin{array}{c}
1\\
i
\end{array}\right)\label{eq:u_XY}
\end{equation}
where $z_{i}$ denotes the position of the input plane. While propagating
through the sample, the $X'$ and $Y'$ components of the probe light
accumulate a relative phase $\delta$, such that at the output plane
$z_{o}$
\begin{flalign}
\mathbf{u}_{X'Y'}(z_{o})= & \left(\begin{array}{cc}
1/\sqrt{2} & 0\\
0 & e^{i\delta}/\sqrt{2}
\end{array}\right)\left(\begin{array}{c}
1\\
i
\end{array}\right)\label{eq:relative-phase}
\end{flalign}
where the sign of $\delta$ depends on the stage of molecular dynamics
(alignment/antialignment). During the alignment stage, $n_{X'}>n_{Y'}$
making $X'$-axis the slow axis, accordingly, the $Y'$ component
leads the $X'$ component and, in the chosen phase convention, $\delta<0$.
Transforming back to the $XY$ basis is achieved by $\mathbf{u}_{XY}(z=z_{o})=R_{Z}(\chi)\mathbf{u}_{X'Y'}(z=z_{o})$,
where $R_{Z}(\chi)$ is the canonical rotation matrix about the $Z$
axis, where $\chi$ is the angle between $X'$, the slow axis of the medium 
(here, the pump polarization axis), and $X$ axes. This results in
\begin{equation}
\mathbf{u}_{XY}(z=z_{o})=\frac{1}{\sqrt{2}}\left(\begin{array}{c}
\cos\chi-ie^{i\delta}\sin\chi\\
ie^{i\delta}\cos\chi+\sin\chi
\end{array}\right).\label{eq:u_XY-after-prop}
\end{equation}
For $\delta<0$ $(\delta>0)$, the Jones vector in Eq. \eqref{eq:u_XY-after-prop}
represents left (right) elliptical light. The major axis of the ellipse
is oriented at angle $\pi/4+\chi$ $(-\pi/4+\chi)$ relative to the
positive $X$ axis (or at angle $\pi/4$ $(-\pi/4)$ relative to
the principal $X'$ axis).

For every point (with azimuth $\varphi$) on the PAF plane in the laboratory frame, the PAF in the $XY$ basis has a specific Jones matrix representation, 
\begin{equation}
M_{\mathrm{R}}=\begin{pmatrix}1 & 0\\
0 & 0
\end{pmatrix}\begin{pmatrix}\cos\varphi & \sin\varphi\\
\sin\varphi & -\cos\varphi
\end{pmatrix},\label{eq:Jones-radial-polarizer}
\end{equation}
where the right matrix represents a $m=1$ vortex plate \citep{Yamaguchi1989,Stalder1996}
(the first element of the PAF). The transformation induced by the vortex
plate is equivalent to the transformation induced by a half-wave plate
with a continuously varying angle of the fast axis, $\varphi/2$.
The left matrix in Eq. \eqref{eq:Jones-radial-polarizer} represents
a linear polarizer oriented along the $X$ axis (the second element
of the PAF). The intensity as a function of the azimuthal
angle at the image plane is given by
\begin{align}
I(\varphi;\delta,\chi)= & |M_{\mathrm{R}}\mathbf{u}_{XY}(z=z_o)|^{2}\nonumber \\
= & \frac{1}{2}-\frac{1}{2}\sin(\delta)\sin\left[2(\varphi-\chi)\right].\label{eq:intensity-afo-phi}
\end{align}

The case of $\chi=0$ corresponds to the molecular alignment along
the $X$ axis. Since the function in Eq. \eqref{eq:intensity-afo-phi}
depends on the difference $\varphi-\chi$, for $\chi\neq0$, the images
are simply rotated counterclockwise by angle $\chi$. Figure \ref{fig:Homo-polar-plots}
shows polar plots of the intensity in Eq. \eqref{eq:intensity-afo-phi}
for the alignment $(\delta<0)$ and antialignment $(\delta>0)$ stages,
and for two angles $\chi=0,\pi/6$. For $\delta=\mp\pi/2$, the out-coming
probe becomes linearly polarized, such that the eight-shaped intensity
pattern is most emphasized and it is oriented at an angle of $+\pi/4$ to the
slow axis of the birefringent medium. When $\delta$ approaches
zero, the pattern continuously transforms into a unit circle (isotropic
case).

\section{Beam Propagation Simulations Homogeneous case \label{sec:Beam-Prop-Homo}}

\noindent We consider a gas of rigid linear molecules excited by 
nonresonant laser pulse (pump) propagating along the $Z$ axis with linear
or shaped polarization (restricted to the $XY$ plane). Generally,
laser excitation results in time-dependent anisotropy and inhomogeneity
of the molecular gas. 

The gaseous medium is probed by collinearly propagating time-delayed
laser pulses (probe pulses). We neglect the difference between the
group velocities of the pump and probe pulses (the Rayleigh length is short enough). 
Also, we assume that the probe pulse duration is much shorter than 
the time scale of molecular rotation, such that for a fixed probe delay, the probe pulse passes
through effectively a time-independent medium. At the first stage,
we assume that the molecular medium is homogeneous which is equivalent
to assuming that the pump intensity is uniform across the $XY$ plane
and along the $Z$ axis. This corresponds to the case when the waist
radius and the Rayleigh range of the probe beam are smaller than those
of the pump beam, such that the probe passes through a relatively
homogeneous portion of the molecular sample. The propagation of the
probe electric field $\mathbf{E}$ through the anisotropic molecular
gas is modeled using the wave equation
\begin{equation}
\nabla^{2}\mathbf{E}-\frac{1}{c^{2}}\overset{\text{\text{\tiny\ensuremath{\bm{\leftrightarrow}}}}}{\boldsymbol{\varepsilon}}_{r}(\tau)\frac{\partial^{2}\mathbf{E}}{\partial t^{2}}=\mathbf{0},\label{eq:WE}
\end{equation}
where $c$ is the speed of light in vacuum, and $\overset{\text{\text{\tiny\ensuremath{\bm{\leftrightarrow}}}}}{\boldsymbol{\varepsilon}}_{r}$
is the tensor of relative permittivity depending on the probe delay,
$\tau$. Derivation of the more general version of Eq. \eqref{eq:WE}
is summarized in the Appendix \ref{sec:Wave-Equation}. We define
the complex amplitude $\mathbf{U}$ by $\mathbf{E}(x,y,z,t)=\mathbf{U}(x,y,z)\exp\left[i(k_{0}z-\omega t)\right]$,
neglecting the pulse nature of the probe, with $\omega$ and $k_{0}$
being the carrier frequency and the vacuum wave number of the probe
light, respectively. Here, we use the same phase convention as in
\citep{Hecht2016} (see Sec. \ref{sec:Qualitative-Homo}). Applying the
paraxial approximation to Eq. \eqref{eq:WE}, we obtain the simplified
equation describing the propagation of the complex amplitude $\mathbf{U}=(U_{X},U_{Y})$
along the $Z$ axis [see Eq. \eqref{eq:WE-paraxial-final}]
\begin{equation}
\frac{\partial\mathbf{U}}{\partial z}=\frac{i}{2k_{0}}\nabla_{T}^{2}\mathbf{U}+\frac{ik_{0}}{2}(\overset{\text{\text{\tiny\ensuremath{\bm{\leftrightarrow}}}}}{\boldsymbol{\varepsilon}}_{r}-\overset{\text{\text{\tiny\ensuremath{\bm{\leftrightarrow}}}}}{\mathbf{I}})\mathbf{U}.\label{eq:WE-paraxial-}
\end{equation}
Here, $\nabla_{T}^{2}$ is the transverse Laplace operator, $\overset{\text{\text{\tiny\ensuremath{\bm{\leftrightarrow}}}}}{\mathbf{I}}$
is the identity matrix. Under ordinary conditions, the relative permittivity
of a molecular gas is simply related to the ensemble-averaged molecular
polarizability tensor, $\braket{\overset{\text{\text{\tiny\ensuremath{\bm{\leftrightarrow}}}}}{\boldsymbol{\alpha}}}_{\mathrm{lab}}$
\begin{equation}
\overset{\text{\text{\tiny\ensuremath{\bm{\leftrightarrow}}}}}{\boldsymbol{\varepsilon}}_{r}=\overset{\text{\text{\tiny\ensuremath{\bm{\leftrightarrow}}}}}{\mathbf{I}}+\frac{N}{\varepsilon_{0}}\braket{\overset{\text{\text{\tiny\ensuremath{\bm{\leftrightarrow}}}}}{\boldsymbol{\alpha}}}_{\mathrm{lab}},\label{eq:permittivity-alpha}
\end{equation}
where $N$ is the number density of the gas, and $\varepsilon_{0}$
is the vacuum permittivity. All the physical quantities in Eq. \eqref{eq:permittivity-alpha}
are expressed in SI units. In terms of $\braket{\overset{\text{\text{\tiny\ensuremath{\bm{\leftrightarrow}}}}}{\boldsymbol{\alpha}}}_{\mathrm{lab}}$,
Eq. \eqref{eq:WE-paraxial-} reads
\begin{equation}
\frac{\partial\mathbf{U}}{\partial z}=\frac{i}{2k_{0}}\nabla_{T}^{2}\mathbf{U}+i\frac{Nk_{0}}{2\varepsilon_{0}}\braket{\overset{\text{\text{\tiny\ensuremath{\bm{\leftrightarrow}}}}}{\boldsymbol{\alpha}}}_{\mathrm{lab}}\mathbf{U}.\label{eq:WE-paraxial-alpha}
\end{equation}

\noindent Equation \eqref{eq:WE-paraxial-alpha} is solved using standard
numerical tools. 
Let us consider two examples of ensemble dynamics and their visualizations: (A) The transition from alignment to antialignment, and (B) the rotation of the alignment axis.

\subsection{Excitation by linearly polarized pump pulse \label{subsec:Homo-Beam-Prop-Linear-Pulse}}

\noindent As the first example, we consider excitation by a short
femtosecond linearly polarized (along $X$ axis) pump pulse. Such
an excitation induces molecular alignment \citep{Stapelfeldt2003,Fleischer2012}---a
transient confinement of molecular axes along the line defined by
the pump polarization. The relatively strong transient alignment appears
immediately after the excitation and disappears shortly after that
because of the molecular angular velocities dispersion. Due to quantum
revivals \citep{Averbukh1989,Robinett2004}, the transient alignment
periodically recurs with a well defined period (in case of linear
molecules) $T_{r}=1/(2Bc)$, where $B=\hbar/(4\pi Ic$) is the molecular
rotational constant, and $I$ is the moment of inertia (more directly, $T_{r}=2\pi I/\hbar$). There are
also fractional revivals emerging at multiples of $T_{r}/4,\,T_{r}/2,\,\mathrm{etc.}$
The degree of alignment is quantified by the expectation value $\braket{\cos^{2}\theta_{X}}$
where $\theta_{X}$ is the angle between the laboratory $X$ axis
and the molecular axis. 
\begin{figure}[h]
\begin{centering}
\includegraphics[width=8cm]{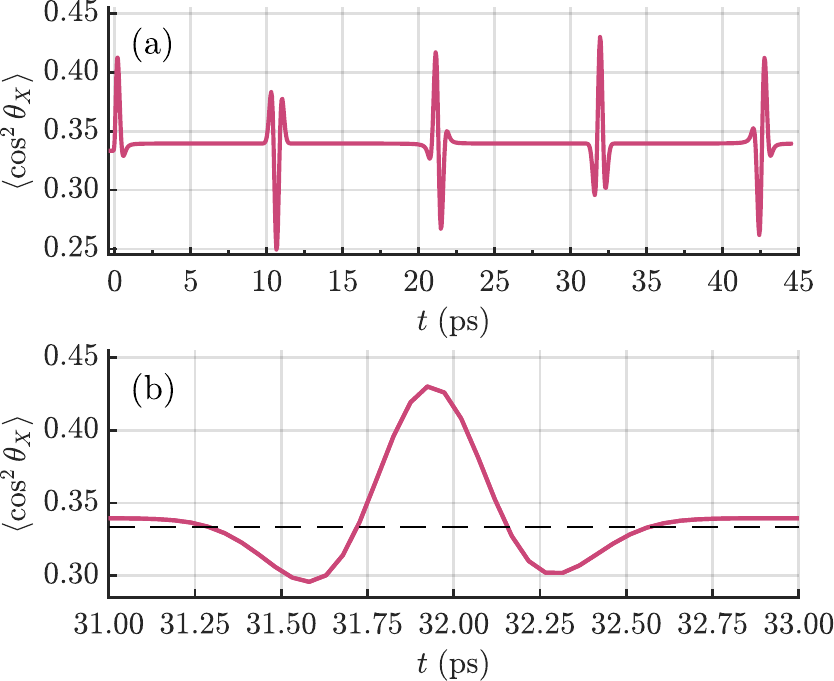}
\par\end{centering}
\caption{(a) Degree of alignment, quantified by $\braket{\cos^{2}\theta_{X}}$.
Pump peak intensity $I_{0}=20\;\mathrm{TW/cm^{2}}$. The full width
at half maximum of the pulse is $100\,\mathrm{fs}$ and the initial
molecular rotational temperature is $T=300\,\mathrm{K}$. (b) Enlarged
portion of panel (a). The dashed line denotes the degree of alignment
in undisturbed gas, $1/3$, which is considered the threshold between alignment to anti-alignment (alignment values lower than $1/3$).\label{fig:Homo-alis-room-temp}}
\end{figure}
 
\begin{figure*}[t]
\begin{centering}
\includegraphics[width=18cm]{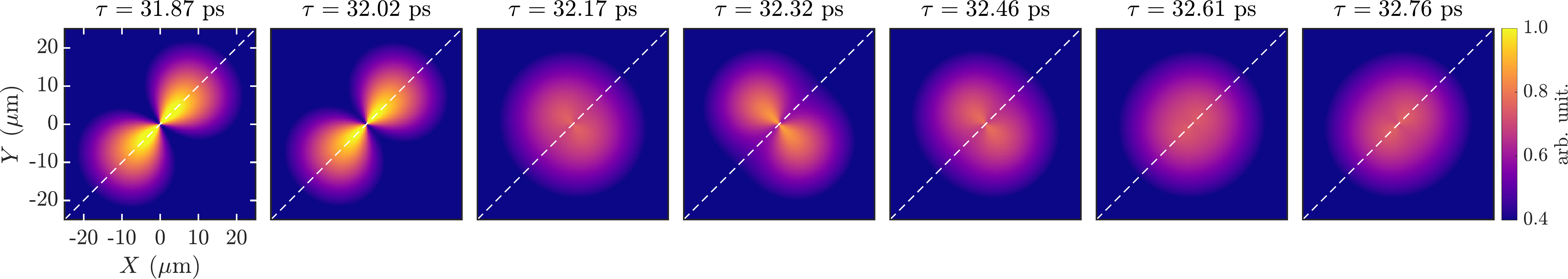}
\par\end{centering}
\caption{Intensity of the probe beam after the PAF (see text) as a function
of the probe delay, $\tau$, during the fractional revival at $t\approx3T_{r}/4$.
Here, the molecules are excited by a linearly polarized along $X$
axis pump pulse. The probe pulse at the input plane is given by Eq.
\eqref{eq:initial_cond}. Probe light propagates through $\approx7\,\mathrm{mm}$
of the molecular gas. The shown images are taken after an additional $1\,\mathrm{mm}$
propagation through an undisturbed gas. \label{fig:Homo-linear-pulse-images}}
\end{figure*}

We consider the alignment of $\mathrm{CO_{2}}$ molecules at initial
rotational temperature $T=300\,\mathrm{K}$. The moment of inertia
of $\mathrm{CO_{2}}$ molecule is $I=280207$ a.u., and the molecular
polarizabilities (in atomic units) along and perpendicular to the
molecular axis are $\alpha_{\parallel}=30.1\,\mathrm{a.u.}$ and $\alpha_{\perp}=14.7\,\mathrm{a.u.}$,
respectively. Figure \ref{fig:Homo-alis-room-temp}(a) shows the degree
of alignment as a function of time (probe delay), following the excitation
at $t=0$.

For quantum mechanical simulations of the expectation values $\braket{\cos^{2}\theta_{X}}$,
the linear molecules were modeled as rigid polarizable rotors. The
interaction energy with nonresonant pump pulse is given by $V=-\mathbf{E}_{\mathrm{pump}}\cdot\mathbf{d}_{\mathrm{ind}}/2=-\mathbf{E}_{\mathrm{pump}}\cdot(\overset{\text{\text{\tiny\ensuremath{\bm{\leftrightarrow}}}}}{\boldsymbol{\alpha}}\mathbf{E}_{\mathrm{pump}})/2,$
where $\mathbf{d}_{\mathrm{ind}}=\overset{\text{\text{\tiny\ensuremath{\bm{\leftrightarrow}}}}}{\boldsymbol{\alpha}}\mathbf{E}_{\mathrm{pump}}$
is the induced dipole, and $\mathbf{E}_{\mathrm{pump}}$ is the pump
electric field. Our numerical scheme for simulating the laser driven
dynamics of complex rigid molecules is described in \citep{Tutunnikov2019}.
In the present case, the scheme is specialized to the case of linear
molecules.

The alignment factor shown in Fig. \ref{fig:Homo-alis-room-temp}
is used to calculate the components of ensemble averaged polarizability
tensor in the laboratory frame \citep{Faucher2011}, $\braket{\overset{\text{\text{\tiny\ensuremath{\bm{\leftrightarrow}}}}}{\boldsymbol{\alpha}}}_{\mathrm{lab}}$
\begin{align}
\braket{\overset{\text{\text{\tiny\ensuremath{\bm{\leftrightarrow}}}}}{\boldsymbol{\alpha}}}_{\mathrm{lab},XX} & (\tau)=\alpha_{\perp}+\Delta\alpha\braket{\cos^{2}\theta_{X}}(\tau),\label{eq:alpha_XX}\\
\braket{\overset{\text{\text{\tiny\ensuremath{\bm{\leftrightarrow}}}}}{\boldsymbol{\alpha}}}_{\mathrm{lab},YY} & (\tau)=\alpha_{\perp}+\frac{\Delta\alpha}{2}\left[1-\braket{\cos^{2}\theta_{X}}(\tau)\right],\label{eq:alpha_YY}
\end{align}
where $\Delta\alpha=\alpha_{\parallel}-\alpha_{\perp}$. In the considered
case, the pump pulse is polarized along the $X$ axis, therefore the
off-diagonal elements of $\braket{\overset{\text{\text{\tiny\ensuremath{\bm{\leftrightarrow}}}}}{\boldsymbol{\alpha}}}_{\mathrm{lab}}$
vanish,
\begin{equation}
\braket{\overset{\text{\text{\tiny\ensuremath{\bm{\leftrightarrow}}}}}{\boldsymbol{\alpha}}}_{\mathrm{lab},XY}=\braket{\overset{\text{\text{\tiny\ensuremath{\bm{\leftrightarrow}}}}}{\boldsymbol{\alpha}}}_{\mathrm{lab},YX}=0\label{eq:off-diagonal-alpha}
\end{equation}

The polarizability tensor components in Eqs. \eqref{eq:alpha_XX}
and \eqref{eq:alpha_YY} are used in Eq. \eqref{eq:WE-paraxial-alpha}
to simulate the probe propagation through the molecular gas. As described
in Sec. \ref{sec:Qualitative-Homo}, the probe light is initially
circularly polarized. The parameters we used are a Gaussian beam with a wavelength $\lambda_\mathrm{probe}=400\,\text{nm}$, waist $\omega_{0,\mathrm{probe}}=20\,\mu\text{m}$, and Rayleigh length $z_{R,\mathrm{probe}}=3.14\,\mathrm{mm}$.

A typical length (along the $Z$ axis) of the effectively excited
molecular medium is about $7\,\mathrm{mm}$. Using Eq. \eqref{eq:WE-paraxial-alpha},
we propagate the probe beam starting at the input plane $z=z_{i}=-3.5\,\mathrm{mm}$
up to the output plane $z=z_{o}=3.5\,\mathrm{mm}$. The initial beam
radius is
\begin{align}
w(z_{i}) & =w_{0,\mathrm{probe}}\sqrt{1+\left(\frac{z_{i}}{z_{R,\mathrm{probe}}}\right)^{2}}=30\,\mathrm{\mu m}.\label{eq:probe-radius}
\end{align}

\begin{figure}
\begin{centering}
\includegraphics[width=7cm]{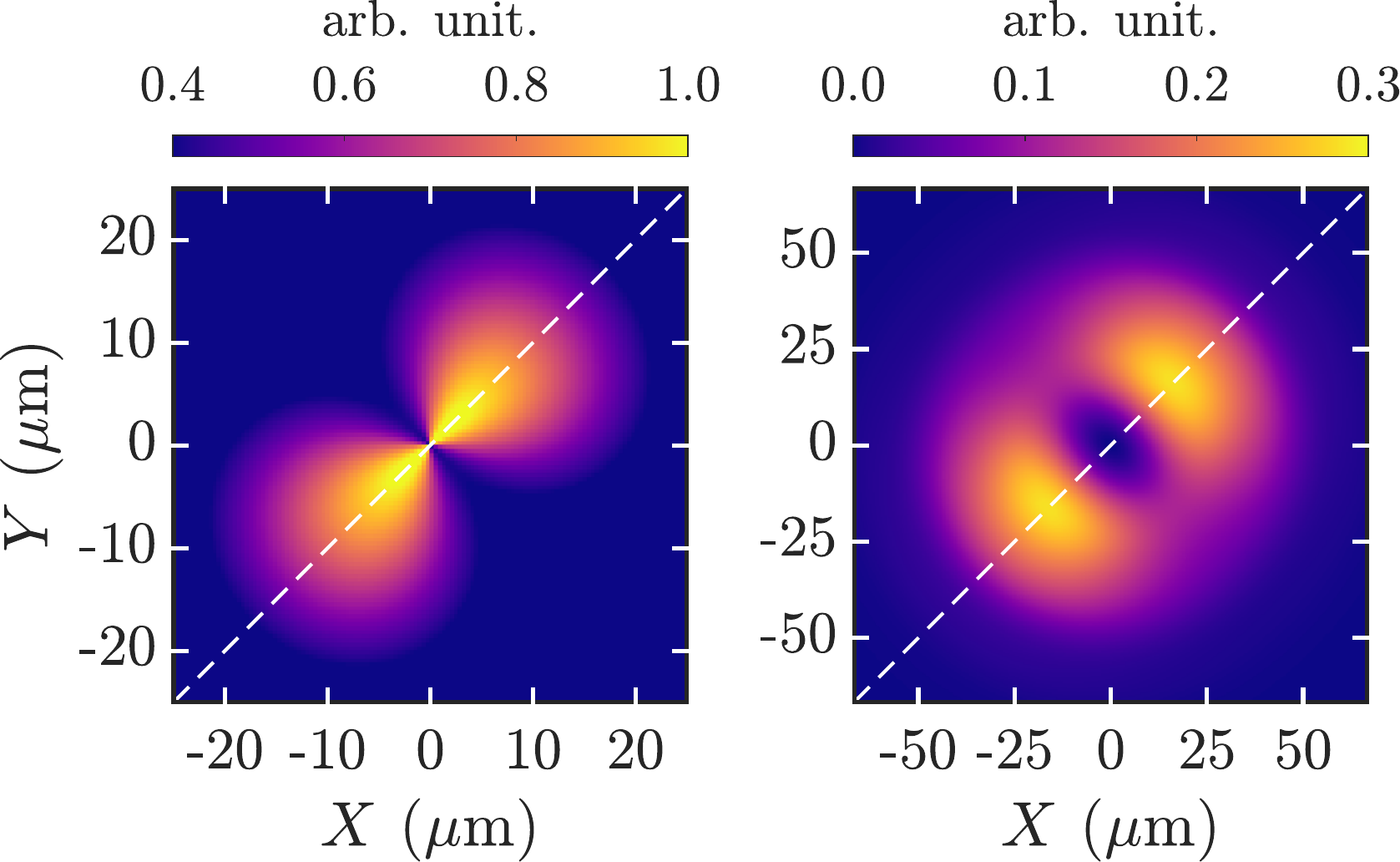}
\par\end{centering}
\caption{Comparison between the images obtained after the PAF.
Left - assuming the vortex plate is attached to the linear polarizer.
Right - when the vortex plate and the linear polarizer are spaced
apart by $2\,\mathrm{mm}$. Here, the probe delay is $\tau=32.02$ (see Fig. \ref{fig:Homo-linear-pulse-images}), and the vortex plate is at a distance of $1\,\mathrm{mm}$ away from the medium.
\label{fig:Homo-beams-comparison}}
\end{figure}

The radius of curvature of the Gaussian beam at the input plane is
\begin{flalign}
R(z_{i}) & =z_{i}\left[1+\left(\frac{z_{R,\mathrm{probe}}}{z_{i}}\right)^{2}\right]=-6.32\,\mathrm{mm},\label{eq:probe-curvature}
\end{flalign}

while the initial complex amplitude is
\begin{equation}
\begin{aligned}U_{X}(x,y,z_{i}) & =\exp\left(-\frac{x^{2}+y^{2}}{w^{2}(z_i)}+ik_0\frac{x^{2}+y^{2}}{2R(z_i)}\right)\\
U_{Y}(x,y,z_{i}) & =iU_{X},
\end{aligned}
\label{eq:initial_cond}
\end{equation}
where $w(z_i)$ and $R(z_i)$ are defined in Eqs. \eqref{eq:probe-radius} and
\eqref{eq:probe-curvature}, respectively. The complex amplitude describing
the probe beam at $z_{o}$ is found by numerically solving the equation
in Eq. \eqref{eq:WE-paraxial-alpha} with the initial condition in
Eq. \eqref{eq:initial_cond}. Here, $X$ and $Y$ components of the
complex amplitude vector are decoupled, because the off-diagonal elements
of the polarizability in the lab frame vanish [see Eq. \eqref{eq:off-diagonal-alpha}].
Exiting the gas sample, the probe passes through the PAF
(see Sec. \ref{sec:Qualitative-Homo}). 

Figure \ref{fig:Homo-linear-pulse-images} shows a series of intensity
maps of the probe beam after the PAF at several delays
during the fractional revival at $t=3T_{r}/4$ (see Fig. \ref{fig:Homo-alis-room-temp}), which is characterized by a transition from molecular alignment to antialignment.
Here, the number density of the gas is set to $N=0.02504\times10^{26}\,\mathrm{m}^{-3}$
[see Eq. \eqref{eq:WE-paraxial-alpha}], corresponding to a pressure
of approximately $0.1\,\mathrm{atm}$. During the alignment stage,
$t\approx31.75-32.12\,\mathrm{ps}$ [see Fig. \ref{fig:Homo-alis-room-temp}(b)],
the intensity pattern is at angle $+\pi/4$ relative to the $X$ axis
(axis of pump polarization, principal optical axis $X'$). During
the antialignment stage, $t\approx31.25-31.75\,\mathrm{ps}$ and $t\approx32.12-32.60\,\mathrm{ps}$
[see Fig. \ref{fig:Homo-alis-room-temp}(b)], the intensity pattern
is at angle $-\pi/4$ relative to the $X$ axis. These results are
in line with the qualitative discussion in Section \ref{sec:Qualitative-Homo}.
Notice, between the revivals, there is a small persistent alignment
which is reflected in weak anisotropy visible in the last panel
in Fig. \ref{fig:Homo-linear-pulse-images}.

Let us note that in practice, the vortex plate and linear
polarizer (forming together the PAF) are spaced apart,
and the vortex plate applies an angle-dependent phase mask [see Eq. \eqref{eq:Jones-radial-polarizer}]
to the incident beam. Therefore, while propagating between the vortex
plate to the polarizer, the incident Gaussian probe beam disperses and evolves into
a doughnut-like shape, such that the experimental images look more
like shown in the right panel of Fig. \ref{fig:Homo-beams-comparison}
(also, see the experimental results in \citep{Bert2019}). Moreover,
in experiments, for practical reasons, the image plane is located much farther from the excitation
volume (far-field measurement). The images shown here do not represent
the actual scale expected in the experiments. The scale depends on
the optical distance and the lenses system between the gas cell
and the imaging device.

\subsection{Excitation by polarization-twisted pump pulse \label{subsec:Homo-Beam-Prop-Twisted-Pulse}}

\begin{figure}[h]
\begin{centering}
\includegraphics[width=7cm]{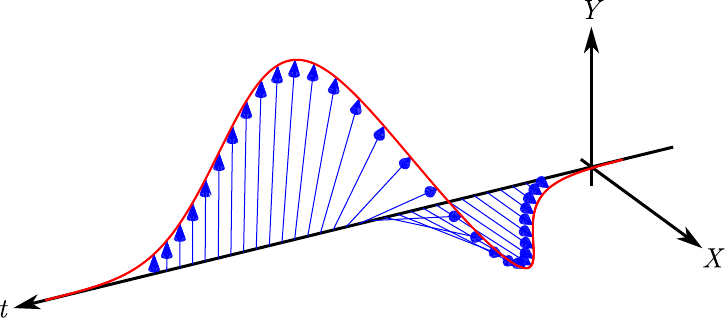}
\par\end{centering}
\caption{Illustration of the polarization twisting, see Eq. \eqref{eq:E-polarization-shaped}.
\label{fig:twisted-pulse-illustration}}
\end{figure}

\noindent As an additional example, we consider the excitation by
a polarization shaped pulse with twisted polarization \citep{Karras2015}.
A polarization-twisted pulse is modeled as a pair of two in-phase overlapping orthogonally
polarized laser pulses with a delay $\tau_{p}$ between them. The
corresponding electric field reads
\begin{equation}
\mathbf{E}_{\mathrm{pump}}(t)=E_{0}[f(t)\mathbf{e}_{X}+f(t-\tau_{p})\mathbf{e}_{Y}]\cos(\omega t),\label{eq:E-polarization-shaped}
\end{equation}
where $E_{0}$ is the peak amplitude, $\mathbf{e}_{X}$ and $\mathbf{e}_{Y}$
are the unit vectors along the laboratory $X$ and $Y$ axes. The
envelope $f(t)$ of each constituent linearly polarized pulse is Gaussian.
Figure \ref{fig:twisted-pulse-illustration} shows an illustration
of the polarization vector twisting from the $X$ axis to the $Y$ axis. 

\begin{figure}[h]
\begin{centering}
\includegraphics[width=8cm]{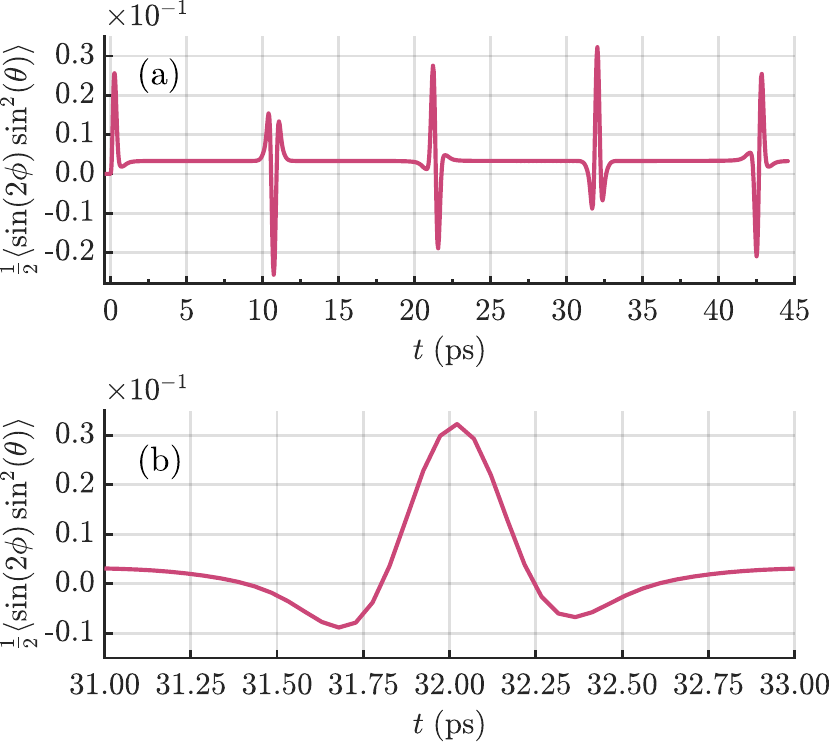}
\par\end{centering}
\caption{(a) The curve is proportional to the off-diagonal element of the polarizability
tensor, see Eq. \eqref{eq:alpha_XY}. Parameters of the polarization
twisted pump pulse: peak intensity $I_{0}=20\;\mathrm{TW/cm^{2}}$,
full width at half maximum of each constituent linearly polarized
pulse is $100\,\mathrm{fs}$, the delay between the pulses is $\tau_{p}=150\,\mathrm{fs}$.
Initial molecular rotational temperature is $T=300\,\mathrm{K}$.
(b) Enlarged portion of panel (a). \label{fig:Homo-off-diag}}
\end{figure}

In contrast to the excitation by a linearly polarized pump
pulse, here, the $Z$ projection of molecular angular momentum is
not conserved as the twisting polarization induces molecular unidirectional
rotation (UDR) about the $Z$ axis \citep{Fleischer2009}. In this
case, the description of molecular rotations required two dynamical
degrees of freedom. It is convenient to use the two standard spherical
coordinate system angles: polar angle $\theta$ between the $Z$ axis
and the molecular axis, and azimuthal angle $\phi$ between the projection
of the molecular axis on the $XY$ plane and the positive $X$ axis.
In the case of UDR the polarizability tensor, $\braket{\overset{\text{\text{\tiny\ensuremath{\bm{\leftrightarrow}}}}}{\boldsymbol{\alpha}}}_{\mathrm{lab}}$
is no longer diagonal, because the principal optical axes ($X'$ and
$Y'$) do not generally overlap with $X$ and $Y$ axes, and their
orientation changes with time. The off-diagonal elements of $\braket{\overset{\text{\text{\tiny\ensuremath{\bm{\leftrightarrow}}}}}{\boldsymbol{\alpha}}}_{\mathrm{lab}}$,
reflecting the unidirectional rotation of the molecular axis \citep{Steinitz2014,Karras2015},
read 
\begin{equation}
\braket{\overset{\text{\text{\tiny\ensuremath{\bm{\leftrightarrow}}}}}{\boldsymbol{\alpha}}}_{\mathrm{lab},XY}=\braket{\overset{\text{\text{\tiny\ensuremath{\bm{\leftrightarrow}}}}}{\boldsymbol{\alpha}}}_{\mathrm{lab},YX}=\frac{\Delta\alpha}{2}\braket{\sin(2\phi)\sin^{2}(\theta)}.\label{eq:alpha_XY}
\end{equation}
Explicit expressions for all the elements of $\braket{\overset{\text{\text{\tiny\ensuremath{\bm{\leftrightarrow}}}}}{\boldsymbol{\alpha}}}_{\mathrm{lab}}$
parameterized by $\theta$ and $\phi$ are given in Appendix \ref{sec:Rotational-Dynamics},
see Eqs. \eqref{eq:App-A}-\eqref{eq:App-D}. Figure \ref{fig:Homo-off-diag}(a)
shows the time dependence of the off diagonal element (divided by
$\Delta\alpha$) in Eq. \eqref{eq:alpha_XY}. Figure \ref{fig:Homo-off-diag}(b)
shows the magnified portion during the fractional quantum revival
at $t\approx3T_{r}/4$.

\noindent 

\begin{figure*}
\begin{centering}
\includegraphics[width=18cm]{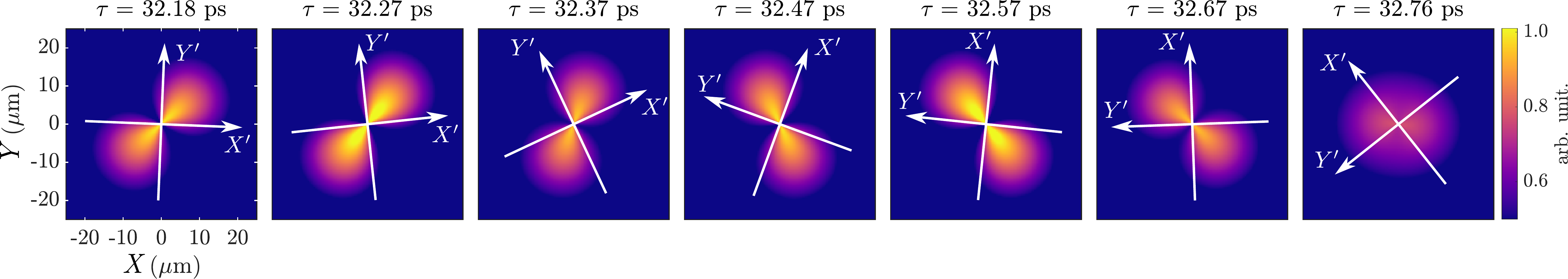}
\par\end{centering}
\caption{Intensity of the probe beam after the PAF as a function
of the probe delay, $\tau$, during the fractional revival at $t\approx3T_{r}/4$.
Here, the molecules are excited by a polarization twisted pulse [see
Eq. \eqref{eq:E-polarization-shaped}]. The delay $\tau$ is measured
from the center of the first linearly polarized (along $X$) pulse
forming the polarization twisted pulse. The rest is the same as in
Fig. \ref{fig:Homo-linear-pulse-images}. \label{fig:Homo-UDR-images}}
\end{figure*}

Figure \ref{fig:Homo-UDR-images} shows a series of intensity images
of the probe beam after the PAF at several delays during
the fractional revival at $t\approx3T_{r}/4$. In addition to the
anisotropy (like in the case of excitation by a linearly polarized
pulse) the principal optical axes, $X'$ and $Y'$, also rotate in
the $XY$ plane. This rotation is the manifestation of molecular UDR.

\noindent 

\section{Qualitative description -- Inhomogeneous Case \label{sec:Qualitative-Inhomo}}

So far, we considered the case of uniformly excited molecular medium,
and only the birefringence effect has been taken into account. Generally,
however, when the probe beam waist is comparable with the waist
 of the pump pulse, the inhomogeneity needs to be taken into
account. The aforementioned inhomogeneity stems from the fact that the pump
intensity decreases away from the optical axis,
resulting in the variation of the molecular alignment across the transverse
plane. In the same manner, on the $Z$ axis, the pump intensity (and its induced molecular effect) reduces with the distance from the focal plane. 

Once again, for simplicity, we consider the case of excitation by
a linearly polarized pump pulse, polarized at an angle $\chi$ to
the $X$ axis. The transverse molecular alignment inhomogeneity acts as an effective
lens, as it manifests in a graded refractive index. Generally, the two orthogonal probe components, $X'$ and $Y'$,
focus differently. During the alignment stage, the $X'$ components
is focused, while the $Y'$ component is defocused and vice versa
during the antialignment stage \citep{Renard2005}. Qualitatively,
the focusing effect can be described by introducing the attenuation
parameter $0\leq a\leq1$, such that [compare to Eq. \eqref{eq:relative-phase}]
\begin{align}
\mathbf{u}_{X'Y'}(z_{o}) & =\left(\begin{array}{cc}
\sqrt{1-a^{2}} & 0\\
0 & ae^{i\delta}
\end{array}\right)\left(\begin{array}{c}
1\\
i
\end{array}\right).\label{eq:relative-phase-a}
\end{align}
For $a=1/\sqrt{2}$, Eq. \eqref{eq:relative-phase-a} reduces to Eq.
\eqref{eq:relative-phase}. Notice, the definition
of parameter $a$ implies that the beam's energy at a fixed radial
distance is conserved. In reality, however, the \emph{total
energy} is conserved. Therefore, $a$ should be treated as an effective parameter accounting for 
the unequal intensities of the two polarization components of the
probe beam at a particular radial distance from the optical axis and at particular $z_o$. The intensity at the output plane
reads [compare with Eq. \eqref{eq:intensity-afo-phi}]
\begin{figure}
\includegraphics{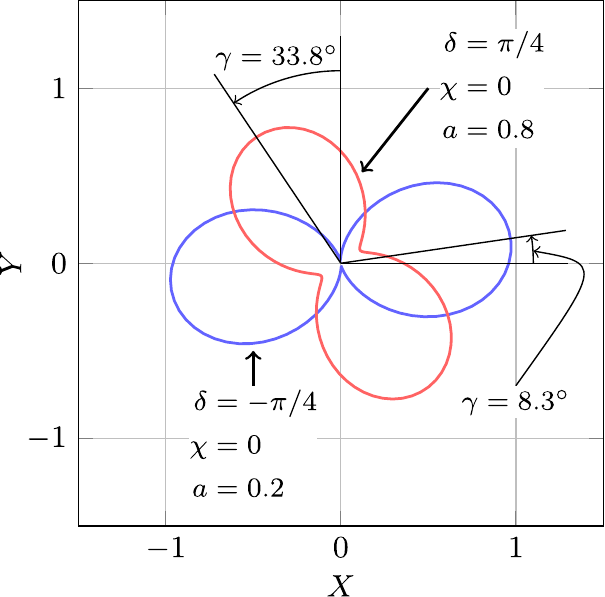}

\caption{Polar plots of the intensity in Eq. \eqref{eq:intensity-afo-phi-left} calculated for alignment (blue) and antialignment (red) induced by a pump polarized along the $X$ axis.
The angle $\gamma$ is given by Eq. \eqref{eq:angle-gamma}. \label{fig:Inhomo-polar-plot-1}}
\end{figure}
\begin{align}
I(\varphi;\delta,\chi,a) & =\frac{1}{2}-a\sqrt{1-a^{2}}\sin(\delta)\sin[2(\varphi-\chi)]\nonumber \\
 & +\left(\frac{1}{2}-a^{2}\right)\cos[2(\varphi-\chi)].\label{eq:intensity-afo-phi-left}
\end{align}

\begin{figure}
\includegraphics{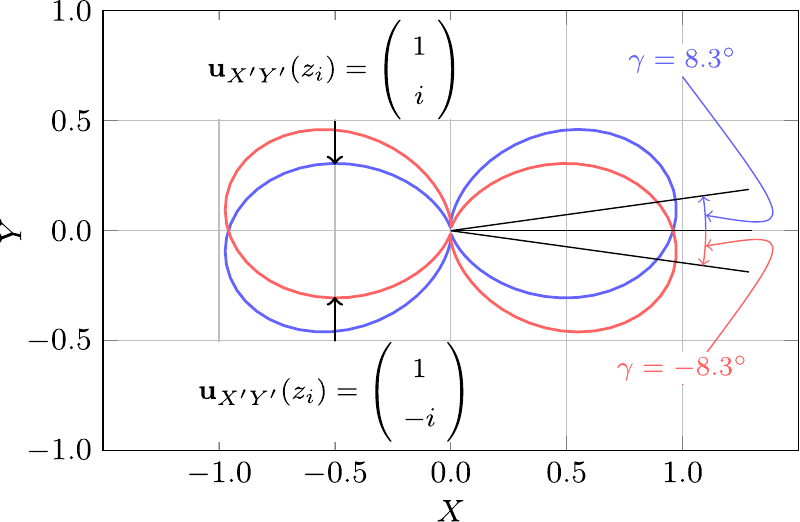}

\caption{Polar plots for left- (blue) and righ-handed (red) probes [see Eqs.
\eqref{eq:intensity-afo-phi-left} and \eqref{eq:intensity-afo-phi-right}].
The angle $\gamma$ for the case of left circular probe is given by
Eq. \eqref{eq:angle-gamma}. The other parameters are $\delta=-\pi/4,\,\chi=0$, and $a=0.2$.
\label{fig:Inhomo-polar-plot-2}}
\end{figure}

Figure \ref{fig:Inhomo-polar-plot-1} shows polar plots of the intensity
in Eq. \eqref{eq:intensity-afo-phi-left}. During the alignment stage
$(\delta<0,\,a<1/\sqrt{2})$, the probe component polarized along
the $X'$ axis is focused, while the $Y'$ components is defocused.
The eight-shaped intensity pattern is effectively pushed closer to
the $X'$ axis. During the antialignment stage $(\delta>0,\,a>1/\sqrt{2})$,
the probe component polarized along the $Y'$ axis is focused, while
the $X'$ components is defocused. The eight-shaped intensity pattern
is effectively pushed closer to the $Y'$ axis. 

The angle $\gamma$ (in radians) between the $X'$ ($Y'$) axis for
the case $\delta<0,\,a<1/\sqrt{2}$ $(\delta>0,\,a>1/\sqrt{2})$ and
the long axis of the eight-shaped intensity pattern (see Fig. \ref{fig:Inhomo-polar-plot-1})
can be found from $\partial_{\varphi}I(\varphi;\delta,\chi,a)|_{\gamma}=0$.
Explicitly, 
\begin{equation}
\gamma=\frac{1}{2}\arctan\left(\frac{2a\sin(\delta)\sqrt{1-a^{2}}}{2a^{2}-1}\right),\label{eq:angle-gamma}
\end{equation}
Generally, angle $\gamma$ also depends on the radial distance from the optical axis and the position along the optical axis. The reason is that the ratio of intensities of the two polarization components changes with radial and longitudinal positions. In the case of negligible focusing, $a\rightarrow1/\sqrt{2}$,
$\gamma\rightarrow\pi/4$ [see Eq. \eqref{eq:intensity-afo-phi}
and Fig. \ref{fig:Homo-polar-plots}].

At this point, we would like to draw the reader's attention to that
in the case of non-negligible focusing, the orientation of the optical
principal axes is ambiguous. Indeed, consider e.g., the blue curve in Fig.
\ref{fig:Inhomo-polar-plot-1}. What is the reason behind $\gamma\neq\pm\pi/4$?
There are two possible reasons (or a combination thereof): (i) the
principal optical axes are rotated $\chi\neq0$, while $\gamma=\pm\pi/4$,
(ii) $X'$ and $Y'$ axes coincide with $X$ and $Y$ axes ($\chi=0$),
but there is a non-negligible focusing, $\gamma\neq\pm\pi/4$. One
way to resolve the ambiguity is to perform an additional measurement
using circular probe light of opposite handedness. It can be shown
that for the right circular probe, the intensity after the PAF
 reads [compare with Eq. \eqref{eq:intensity-afo-phi-left}]
\begin{align}
I(\varphi;\delta,\chi,a) & =\frac{1}{2}+a\sqrt{1-a^{2}}\sin(\delta)\sin[2(\varphi-\chi)]\nonumber \\
 & +\left(\frac{1}{2}-a^{2}\right)\cos[2(\varphi-\chi)].\label{eq:intensity-afo-phi-right}
\end{align}
Notice the plus sign in front of the second terms compared to Eq.
\eqref{eq:intensity-afo-phi-left}. Moreover, angle $\gamma$
[see Eq. \eqref{eq:angle-gamma}] has an opposite sign for the
opposite circular polarizations. Figure \ref{fig:Inhomo-polar-plot-2}
compares the intensities obtained for the probes of opposite handedness
in the case of non-negligible focusing.

The pictures obtained from the two measurements together, in principle,
enable to determine the orientations of the optical principal
axes (and to assess the focusing strength). As shown in Fig. \ref{fig:Inhomo-polar-plot-2}, the principal
axis bisects the angle between the long axis of the two intensity
patterns.

\section{Beam Propagation Simulations -- Inhomogeneous case \label{sec:Beam-Propg-Inhomo}}

\noindent In the case of non-negligible inhomogeneity, the permittivity
of the medium becomes coordinate-dependent and the wave equation reads
[compare with Eq. \eqref{eq:WE}] 
\begin{equation}
\nabla^{2}\mathbf{E}-\frac{1}{c^{2}}\overset{\text{\text{\tiny\ensuremath{\bm{\leftrightarrow}}}}}{\boldsymbol{\varepsilon}}_{r}(x,y,z;\tau)\frac{\partial^{2}\mathbf{E}}{\partial t^{2}}=0.\label{eq:WE-inhomogeneous}
\end{equation}
Defining the complex amplitude $\mathbf{U}$, and using the paraxial
approximation, we arrive at the same equation as Eq. \eqref{eq:WE-paraxial-alpha},
but with coordinate dependent polarizability tensor $\braket{\overset{\text{\text{\tiny\ensuremath{\bm{\leftrightarrow}}}}}{\boldsymbol{\alpha}}}_{\mathrm{lab}}$.
The derivation is summarized in Appendix \ref{sec:Wave-Equation}.
Again, we consider the dynamics of alignment to antialignment transition, and the rotation of the alignment axis.

\subsection{Excitation by linearly polarized pump pulse \label{subsec:Inhomo-Beam-Prop-Linear-Pulse}}

Considering first the excitation by a linearly polarized along the
$X$ axis pump pulse. Similar to Subsec. \ref{subsec:Homo-Beam-Prop-Linear-Pulse},
the polarizability $\braket{\overset{\text{\text{\tiny\ensuremath{\bm{\leftrightarrow}}}}}{\boldsymbol{\alpha}}}_{\mathrm{lab}}$
is diagonal, and its components are given by Eqs. \eqref{eq:alpha_XX}
and \eqref{eq:alpha_YY}. However, in contrast to Subsec. \ref{subsec:Homo-Beam-Prop-Linear-Pulse},
the degree of alignment now also depends on the coordinates $\braket{\cos^{2}\theta_{X}}=\braket{\cos^{2}\theta_{X}}(x,y,z;\tau)$.
Following \citep{Bert2019}, we assume that the probe light is derived
from the pump light by frequency doubling, and both have Gaussian
profiles. Table \ref{tab:beams-parameters-2} summarize the beams
parameters used. To simplify matters, we neglect the inhomogeneity
along the propagation direction ($Z$ axis), such that $\braket{\cos^{2}\theta_{X}}=\braket{\cos^{2}\theta_{X}}(x,y;\tau)$. 

\begin{table}[h]
\begin{centering}
\begin{tabular}{c|c|c|c}
Beam & $\lambda\,(\mathrm{nm})$ & $w_{0}\,(\mathrm{\mu m})$ & $z_{R}\,(\mathrm{mm})$\tabularnewline
\hline 
Pump & $800$ & $30$ & $3.53$\tabularnewline
\hline 
Probe & $400$ & $20$ & $3.14$\tabularnewline
\end{tabular}
\par\end{centering}
\caption{Summary of the beams parameters. Here $\lambda$, $w_{0}$, $z_{R}$
are vacuum wavelength, waist radius and Rayleigh range. \label{tab:beams-parameters-2}}
\end{table}

\begin{figure}
\begin{centering}
\includegraphics[width=8cm]{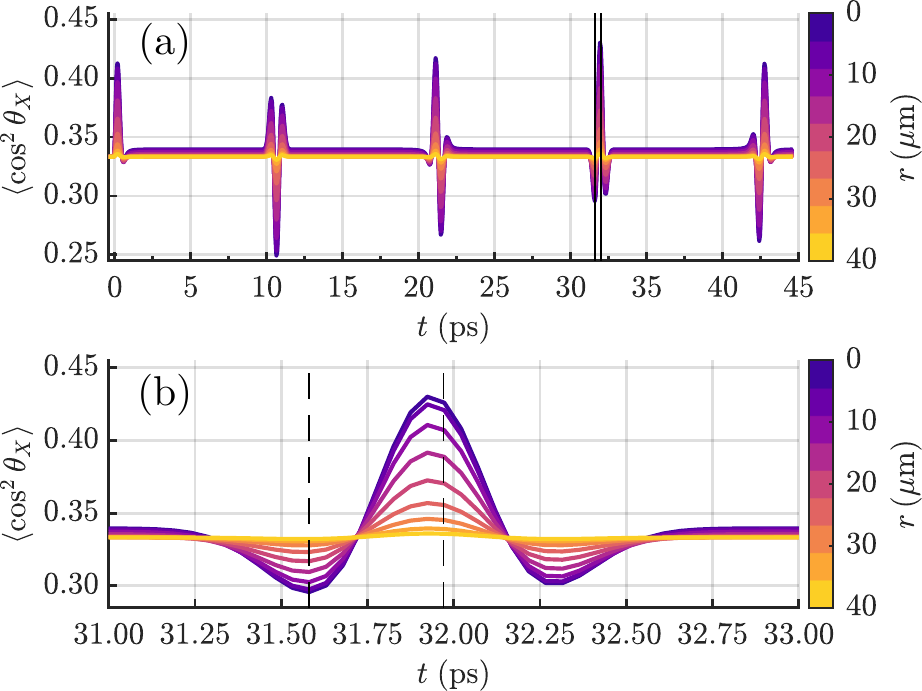}
\par\end{centering}
\caption{(a) Degree of alignment at various transverse distances $r$ from
the optical axis. The pump peak intensity is $I(r=0)=20\;\mathrm{TW/cm^{2}}$
and the pump pulse duration at full width at half maximum (FWHM) is $100\,\mathrm{fs}$. Initial molecular rotational
temperature is $T=300\,\mathrm{K}$. (b) Enlarged portion of panel
(a). Black vertical lines denote the delays shown in Fig. \ref{fig:Inhomo-alis-XY}.
\label{fig:Inhomo-alis}}
\end{figure}

Figure \ref{fig:Inhomo-alis} shows several curves of time-dependent alignment
factors at various transverse distances, $r$ from the optical axis
for the case of $\mathrm{CO_{2}}$ molecules. Since the intensity
of the pump decreases with the distance from the optical axis, the
overall degree of alignment decreases as well. Here, the initial rotational
temperature is $T=300\,\mathrm{K}$. 

\begin{figure}
\begin{centering}
\includegraphics[width=8cm]{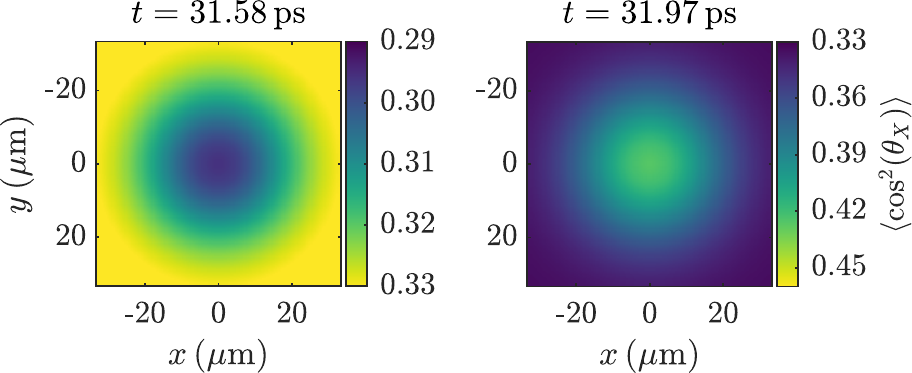}
\par\end{centering}
\caption{The spatial variation of the degree of alignment at two times denoted
by vertical lines in Fig. \ref{fig:Inhomo-alis}. \label{fig:Inhomo-alis-XY}}
\end{figure}

Figure \ref{fig:Inhomo-alis-XY} shows the spatial dependence of the
degree of alignment in the $XY$ plane. The left panel $(t=31.58\,\mathrm{ps})$
corresponds to the antialignemnt stage, while the right panel $(t=31.97\,\mathrm{ps})$
corresponds to the alignment stage. These two moments are denoted
by the vertical lines in Fig. \ref{fig:Inhomo-alis}. Considering
Eqs. \eqref{eq:alpha_XX} and \eqref{eq:alpha_YY} and the wave equation
in Eq. \eqref{eq:WE-paraxial-alpha}, the $X'$ and $Y'$ components
(here, same as $X$ and $Y$) of the probe pulse pass through effective
lenses with curvatures $[\braket{\cos^{2}\theta_{X}}(x,y)-1/3]$ and $[\braket{\cos^{2}\theta_{Y}}(x,y)-1/3]$ [with $\braket{\cos^{2}\theta_{Y}}(x,y)=-2\braket{\cos^{2}\theta_{X}}(x,y)]$,
respectively (see Fig. \ref{fig:Inhomo-alis-XY}). Accordingly, the
$X'$ component is focused, while the $Y'$ component is defocused,
and vice versa during the antialignment stage.

\begin{figure*}
\begin{centering}
\includegraphics[width=18cm]{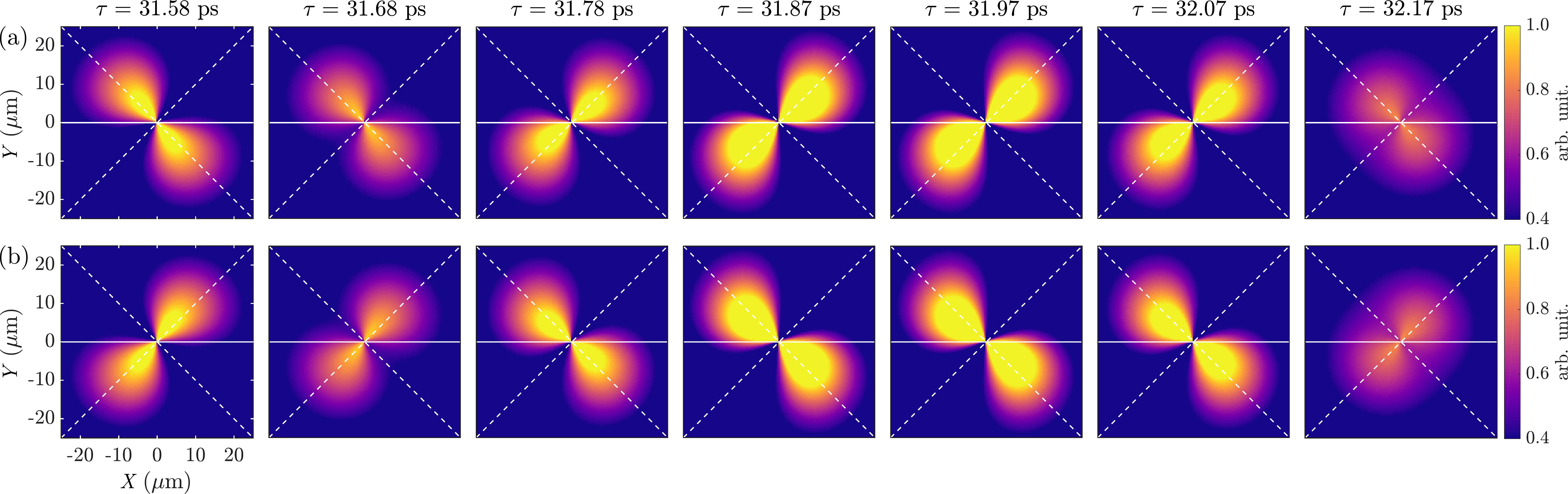}
\par\end{centering}
\caption{Intensity of the probe beam after the PAF as a function
of the delay, $\tau$, during the fractional revival at $t\approx3T_{r}/4$.
Here, the molecules are excited by a linearly polarized along $X$-axis pump pulse. Probe light propagates through $\approx7\,\mathrm{mm}$
of the molecular gas. The beams' parameters are defined in Table \ref{tab:beams-parameters-2}.
The shown images are taken after an additional propagation of $1\,\mathrm{mm}$ through an undisturbed gas.
(a) Left circular probe. (b) Right circular probe. \label{fig:Inhomo-linear-pulse-images}}
\end{figure*}

Figure \ref{fig:Inhomo-linear-pulse-images} shows a series of intensity
images of the probe beam after the PAF at several delays
during the fractional revival at $t=3T_{r}/4$ (see Fig. \ref{fig:Inhomo-alis}).
Here, the number density of the gas is set to $N=0.1252\times10^{26}\,\mathrm{m}^{-3}$
[see Eq. \eqref{eq:WE-paraxial-alpha}], corresponding to pressure
of approximately $0.5\,\mathrm{atm}$. Panel (a) shows the intensities
obtained using a left circular probe [see Eq. \eqref{eq:initial_cond}],
while panel (b) shows the results obtained using a right circular probe
($U_{Y}=-iU_{X}$). In both cases, due to the focusing effect, the
long axes of the intensity patterns slightly deviate from the diagonals. As mentioned previously (see Sec. \ref{sec:Qualitative-Inhomo}), the degree of deviation of the long axis of the eight-shaped intensity from $\pm45^{\circ}$, generally, depends on the radial distance $r$ and the PAF/output plane's position. Here, the radial dependence is barely noticeable, while the change with the longitudinal distance, $z$ is visible (not shown).
The deviation is in the opposite directions for the opposite circular
polarizations. This allows determining the orientations of the principal
optical axes unambiguously. It is important to emphasize that due
to the symmetry of excitation, the orientations of the optical principal
axes, $X'$ and $Y'$, are independent of the distance from the optical
axis.

\subsection{Excitation by polarization-twisted pump pulse \label{subsec:Inhomo-Beam-Prop-Twisted-Pulse}}

\begin{figure}[H]
\begin{centering}
\includegraphics[width=8cm]{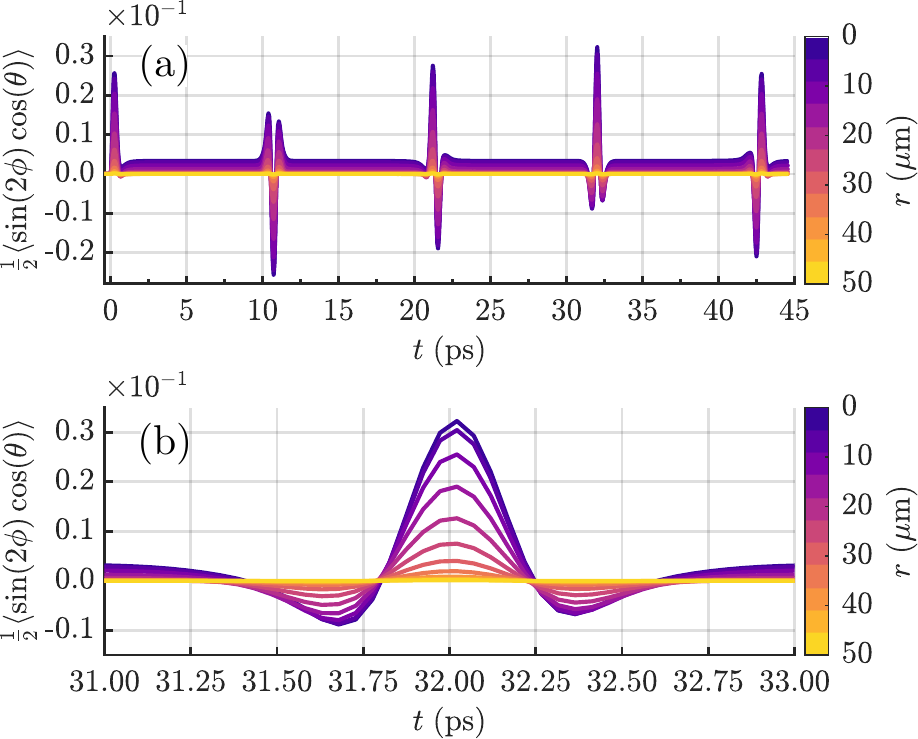}
\par\end{centering}
\caption{The curves are proportional to the off-diagonal element of the polarizability
tensor at various transverse distances, $r$ from the optical axis.
Parameters of the polarization-twisted pump pulse: peak intensity
$I(r=0)=20\;\mathrm{TW/cm^{2}}$ and the 
width (FWHM) of each constituent linearly polarized
pulse is $100\,\mathrm{fs}$, the delay between the pulses is $\tau_{p}=150\,\mathrm{fs}$.
Initial molecular rotational temperature is $T=300\,\mathrm{K}$. (b) Enlarged portion of panel
(a). \label{fig:Inhomo-off-diag}}
\end{figure}

Next, we consider the excitation by the polarization-twisted pulse
[see Eq. \eqref{eq:E-polarization-shaped} and Fig. \ref{fig:twisted-pulse-illustration}]. Similar to Subsec.
\ref{subsec:Homo-Beam-Prop-Twisted-Pulse}, the polarization-twisted
pulse induces molecular UDR, which recurs during the fractional revivals.
Figure \ref{fig:Inhomo-off-diag} shows the off-diagonal element of
the polarizability [divided by $\Delta\alpha$, see Eq. \eqref{eq:alpha_XY}]
for various transverse distances, $r$ from the optical axis. The
existence of non-zero time-dependent (probe delay dependent) off-diagonal
elements of the polarizability imply that the orientation of the optical
principal axes changes with time.

There is an important difference between the excitation by polarization-twisted
pulse having inhomogeneous intensity profile and the excitation by
polarization-twisted pulse with homogeneous intensity considered in
Subsec. \ref{subsec:Homo-Beam-Prop-Twisted-Pulse}. In the case of
twisted pulse the orientation of the optical axes generally
depends on the radial distance from the optical axis. The physical
reason for this is that the efficiency of the induced molecular UDR
depends on the parameters of the polarization-twisted pulses.

\begin{figure*}
\begin{centering}
\includegraphics[width=18cm]{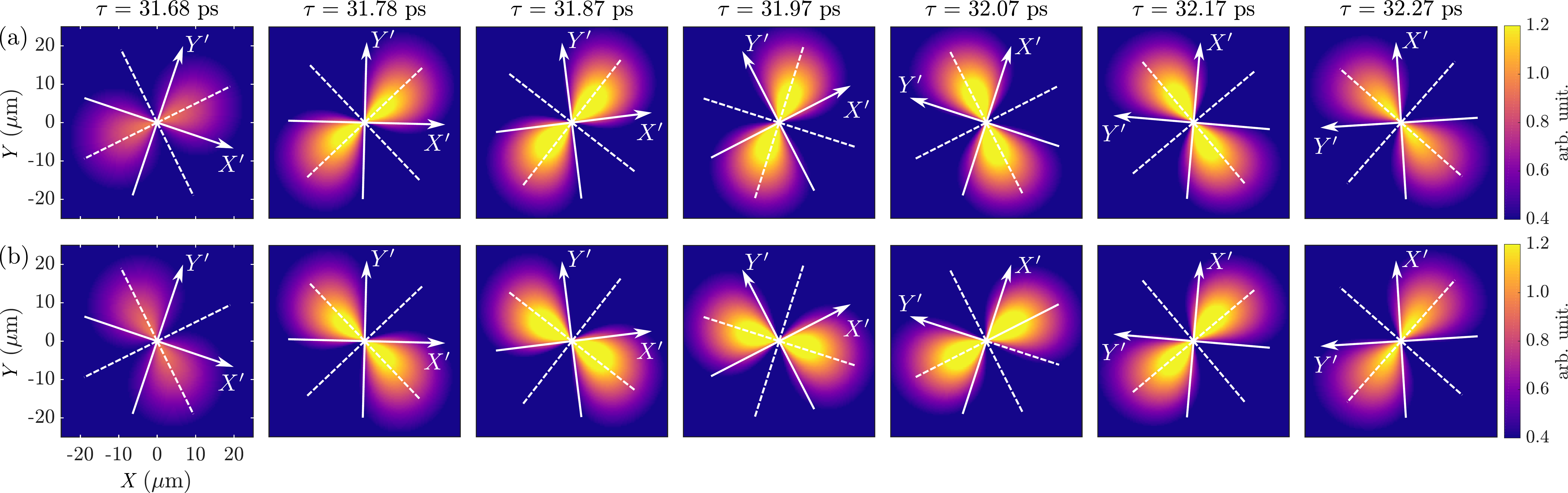}
\par\end{centering}
\caption{Intensity patterns after the PAF as a function of the
probe delay, $\tau$, during the fractional revival at $t\approx3T_{r}/4$.
Here, the molecules are excited by a polarization-twisted pulse [see
Eq. \eqref{eq:E-polarization-shaped}]. The delay is measured from
the peak of the first linearly polarized pulse. Rest of the parameters
are similar to Fig. \ref{fig:Inhomo-linear-pulse-images}. (a) Left
circular probe. (b) Right circular probe. \label{fig:Inhomo-UDR-images}}
\end{figure*}

Figure \ref{fig:Inhomo-UDR-images} shows a series of intensity images
like in Fig. \ref{fig:Inhomo-linear-pulse-images}, but for the case
of excitation by polarization twisted pulse. The white arrows denote
the principal optical axes. As mentioned above, the orientation of
$X'$ and $Y'$ axes depends on the radial distance, however here
the effect remains marginal within the waist of the pump
pulse $(w_{0,\mathrm{pump}}=30\,\mathrm{\mu m})$. The shown principal
axes were found by diagonalizing $\braket{\overset{\text{\text{\tiny\ensuremath{\bm{\leftrightarrow}}}}}{\boldsymbol{\alpha}}}_{\mathrm{lab}}$
at $r=20\,\mathrm{\mu m}$.

Thanks to the weak $r$ dependence, the long axes of the intensity
patterns deviate from the chosen principal axes in approximately the
opposite sense for the probes of opposite circular polarization.

\section{Conclusions \label{sec:Conclusions}}

We theoretically analyzed the optical imaging scheme introduced in \citep{Bert2019}, allowing direct visualization of the instantaneous orientation of the optical principal axes in a gas of laser-excited coherently rotating molecules.
Time-delayed circularly polarized probe pulses are used to investigate the time-dependent optical birefringence of the molecular medium. A polarization axis finder (PAF) is used to visualize the polarization of the probe pulse, which contains the information on the optical principal axes. While in this work, we considered two examples of molecular excitation by short pulses, the same approach applies to the visualization of other complex molecular states created by means of tailored laser pulses, e.g., molecular super rotors \citep{Korobenko2014}.

Future work may generalize the proposed scheme to asymmetric-top molecules, which generally have three distinct polarizability axes.
The scheme may be promising for imaging alignment and unidirectional rotation of complex molecules (including chiral ones). Visualization of the dynamics of such molecules in traditional methods, e.g., Coulomb explosion-based methods, is challenging due to the multitude of possible fragments and ionization channels.
The optical measurement of alignment dynamics may stimulate the development
of imaging schemes suitable for tracing out intricate molecular \emph{orientation} dynamics.

\begin{acknowledgments}
This work was supported by the Israel Science Foundation (Grant No. 746/15), the CNRS, the ERDF Operational Programme-Burgundy, the EIPHI Graduate School (Contract No. ANR-17-EURE-0002), and the Associate (CNRS\&Weizmann) International ImagiNano Laboratory. IA acknowledges support as the Patricia Elman Bildner Professorial Chair. This research was made possible in part by the historic generosity of the Harold Perlman Family. 
\end{acknowledgments}

\appendix

\section{Wave Equation \label{sec:Wave-Equation}}

Here, we derive the wave equation describing the propagation of a
Gaussian beam in a non-magnetic, inhomogeneous, and time independent
molecular gas. Maxwell's equations (SI units) in matter which is free
of currents and charges are
\begin{align}
\nabla\times\mathbf{H} & =\frac{\partial\mathbf{D}}{\partial t},\label{eq:Max1}\\
\nabla\times\mathbf{E} & =-\frac{\partial\mathbf{B}}{\partial t},\label{eq:Max2}\\
\nabla\cdot\mathbf{D} & =0,\label{eq:Max3}\\
\nabla\cdot\mathbf{B} & =0,\label{eq:Max4}
\end{align}
where $\mathbf{E}$ and $\mathbf{B}$ are the electric and magnetic
field vectors, $\mathbf{H}$ and $\mathbf{D}$ are the magnetizing
and displacement field vectors. In addition to the Maxwell's equation,
we use the constitutive relations
\begin{equation}
\mathbf{D}=\varepsilon_{0}\overset{\text{\text{\tiny\ensuremath{\bm{\leftrightarrow}}}}}{\boldsymbol{\varepsilon}}_{r}\mathbf{E}\qquad\mathbf{B}=\mu_{0}\overset{\text{\text{\tiny\ensuremath{\bm{\leftrightarrow}}}}}{\boldsymbol{\mu}}_{r}\mathbf{H},
\end{equation}
where $\varepsilon_{0}$ and $\overset{\text{\text{\tiny\ensuremath{\bm{\leftrightarrow}}}}}{\boldsymbol{\varepsilon}}_{r}$
are the vacuum and relative permittivities, respectively, $\mu_{0}$
and $\overset{\text{\text{\tiny\ensuremath{\bm{\leftrightarrow}}}}}{\boldsymbol{\mu}}_{r}$
are the vacuum and relative permeabilities, respectively. We assume
a non-magnetic medium, $\overset{\text{\text{\tiny\ensuremath{\bm{\leftrightarrow}}}}}{\boldsymbol{\mu}}_{r}=1$.
Relative permittivity $\overset{\text{\text{\tiny\ensuremath{\bm{\leftrightarrow}}}}}{\boldsymbol{\varepsilon}}_{r}$
is position dependent, but time independent tensor. Applying the curl
operator to the Eq. \eqref{eq:Max2} results in
\begin{align}
\nabla\times\left[\left(\nabla\times\mathbf{E}\right)+\frac{\partial\mathbf{B}}{\partial t}\right] & =\nabla\times\left(\nabla\times\mathbf{E}\right)\nonumber \\
+\mu_{0}\frac{\partial}{\partial t}\left(\nabla\times\mathbf{H}\right) & =0,\label{eq:intermediate-WE-1}
\end{align}
where the time derivative and curl operators were interchanged. Substituting
$\nabla\times\mathbf{H}=\partial_{t}\mathbf{D}$ [see Eq. \eqref{eq:Max1}],
we get
\begin{align}
\nabla\times\left(\nabla\times\mathbf{E}\right)+\mu_{0}\frac{\partial^{2}\mathbf{D}}{\partial t^{2}} & =\nabla\times\left(\nabla\times\mathbf{E}\right)\nonumber \\
+\mu_{0}\varepsilon_{0}\overset{\text{\text{\tiny\ensuremath{\bm{\leftrightarrow}}}}}{\boldsymbol{\varepsilon}}_{r}\frac{\partial^{2}\mathbf{E}}{\partial t^{2}} & =0.\label{eq:intermediate-WE-2}
\end{align}

Using the vector identity $\nabla\times\left(\nabla\times\mathbf{A}\right)=\nabla\left(\nabla\cdot\mathbf{A}\right)-\nabla^{2}\mathbf{A}$,
we can simplify Eq. \eqref{eq:intermediate-WE-2}
\begin{equation}
\nabla^{2}\mathbf{E}-\mu_{0}\varepsilon_{0}\overset{\text{\text{\tiny\ensuremath{\bm{\leftrightarrow}}}}}{\boldsymbol{\varepsilon}}_{r}\frac{\partial^{2}\mathbf{E}}{\partial t^{2}}=\nabla\left(\nabla\cdot\mathbf{E}\right).\label{eq:WE-with-RHS}
\end{equation}
For our applications, we would like to simplify Eq. \eqref{eq:WE-with-RHS}
by neglecting $\nabla\left(\nabla\cdot\mathbf{E}\right)$. To estimate
the relative size of $\nabla\left(\nabla\cdot\mathbf{E}\right)$,
we assume that $\overset{\text{\text{\tiny\ensuremath{\bm{\leftrightarrow}}}}}{\boldsymbol{\varepsilon}}_{r}=\boldsymbol{\varepsilon}_{r}(x,y,z)$
is a scalar and use Eq. \eqref{eq:Max3}, $\nabla\cdot\mathbf{D}=\varepsilon_{0}\nabla\cdot(\varepsilon_{r}\mathbf{E})=0$.
This corresponds to the case when the electric field is linearly polarized
along one of the principal axes of $\overset{\text{\text{\tiny\ensuremath{\bm{\leftrightarrow}}}}}{\boldsymbol{\varepsilon}}_{r}$.
Using the vector identity $\nabla\cdot\left(\psi\mathbf{A}\right)=\psi\nabla\cdot\mathbf{A}+(\nabla\psi)\cdot\mathbf{A}$,
we have $\varepsilon_{0}\nabla\cdot(\varepsilon_{r}\mathbf{E})=\varepsilon_{r}\nabla\cdot\mathbf{E}+(\nabla\varepsilon_{r})\cdot\mathbf{E}=0$.
In other words, in case of a scalar relative permittivity the right
hand side of the equation is $-\nabla\{[(\nabla\varepsilon_{r})/\varepsilon_{r}]\cdot\mathbf{E}\}$.
Compared with the second term on the left hand, the right hand side
can be neglected when the relative change in $\varepsilon_{r}$ over
the distance of one wave-length must be much less than unity \citep{Dietrich1989}.
Finally, we obtain the following wave equation \citep{Casperson1973}
\begin{equation}
\nabla^{2}\mathbf{E}-\mu_{0}\varepsilon_{0}\overset{\text{\text{\tiny\ensuremath{\bm{\leftrightarrow}}}}}{\boldsymbol{\varepsilon}}_{r}\frac{\partial^{2}\mathbf{E}}{\partial t^{2}}=0\label{eq:Wave-Equation}
\end{equation}

To further simplify Eq. \eqref{eq:Wave-Equation}, we assume that
the electric field propagates along the $Z$ axis and substitute $\mathbf{E}(x,y,z,t)=\mathbf{U}(x,y,z)\exp\left[i(k_{0}z-\omega t)\right]$. The chosen form of $\mathbf{E}$ implies that we neglect the pulse nature of the light.
Here $\mathbf{U}(x,y,z)=(U_{X},U_{Y},0)$ is the complex amplitude.
For consistency with Section \ref{sec:Qualitative-Homo}, we use Hecht's
phase convention \citep{Hecht2016}, $k_{0}z-\omega t$. The result
reads

\begin{equation}
(\omega^{2}\overset{\text{\text{\tiny\ensuremath{\bm{\leftrightarrow}}}}}{\boldsymbol{\varepsilon}}_{r}-c^{2}k_{0}^{2}\overset{\text{\text{\tiny\ensuremath{\bm{\leftrightarrow}}}}}{\mathbf{I}})\mathbf{U}+c^{2}\left(2ik_{0}\frac{\partial\mathbf{U}}{\partial z}+\nabla^{2}\mathbf{U}\right)=0,\label{eq:WE-before-paraxial}
\end{equation}
where $c=1/\sqrt{\mu_{0}\varepsilon_{0}}$, $\mathbf{I}$ is the identity
matrix, and $\mathbf{e}_{Z}$ is the unit vector along the $Z$ axis.
Next, we make the paraxial approximation and neglect the second derivative
with respect to $z$, such that Eq. \ref{eq:WE-before-paraxial} becomes 

\begin{equation}
(\omega^{2}\overset{\text{\text{\tiny\ensuremath{\bm{\leftrightarrow}}}}}{\boldsymbol{\varepsilon}}_{r}-c^{2}k_{0}^{2}\overset{\text{\text{\tiny\ensuremath{\bm{\leftrightarrow}}}}}{\mathbf{I}})\mathbf{U}+c^{2}\left(2ik_{0}\frac{\partial\mathbf{U}}{\partial z}+\nabla_{T}^{2}\mathbf{U}\right)=0,\label{eq:WE-paraxial}
\end{equation}
where $\nabla_{T}^{2}$ is the transverse Laplacian operator. Rearrangement
yields 
\[
\frac{\partial\mathbf{U}}{\partial z}=\frac{i}{2k_{0}}\nabla_{T}^{2}\mathbf{U}+\frac{ik_{0}}{2}\left(\frac{\omega^{2}}{c^{2}k_{0}^{2}}\overset{\text{\text{\tiny\ensuremath{\bm{\leftrightarrow}}}}}{\boldsymbol{\varepsilon}}_{r}-\overset{\text{\text{\tiny\ensuremath{\bm{\leftrightarrow}}}}}{\mathbf{I}}\right)\mathbf{U}.
\]
Finally, we substitute $\omega^{2}/(c^{2}k_{0}^{2})=1$, such that
\begin{equation}
\frac{\partial\mathbf{U}}{\partial z}=\frac{i}{2k_{0}}\nabla_{T}^{2}\mathbf{U}+\frac{ik_{0}}{2}(\overset{\text{\text{\tiny\ensuremath{\bm{\leftrightarrow}}}}}{\boldsymbol{\varepsilon}}_{r}-\overset{\text{\text{\tiny\ensuremath{\bm{\leftrightarrow}}}}}{\mathbf{I}})\mathbf{U}.\label{eq:WE-paraxial-final}
\end{equation}

\section{Rotational Dynamics \label{sec:Rotational-Dynamics}}

For free rigid linear molecules in the gas phase $\braket{\overset{\text{\text{\tiny\ensuremath{\bm{\leftrightarrow}}}}}{\boldsymbol{\alpha}}}_{\mathrm{lab}}$
is simply a constant, where $\alpha_{\parallel,\perp}$ are the polarizabilities
along and perpendicular to the molecular axis, respectively. In case
of laser excited molecular gas, however $\braket{\overset{\text{\text{\tiny\ensuremath{\bm{\leftrightarrow}}}}}{\boldsymbol{\alpha}}}_{\mathrm{lab}}$
is generally position dependent anisotropic tensor. The relation between
the polarizability expressed in the molecule-fixed frame and the polarizability
expressed in the laboratory-fixed frame
\begin{equation}
\overset{\text{\text{\tiny\ensuremath{\bm{\leftrightarrow}}}}}{\boldsymbol{\alpha}}_{\mathrm{lab}}=\overset{\text{\text{\tiny\ensuremath{\bm{\leftrightarrow}}}}}{\mathbf{R}}^{T}\overset{\text{\text{\tiny\ensuremath{\bm{\leftrightarrow}}}}}{\boldsymbol{\alpha}}_{\mathrm{mol}}\overset{\text{\text{\tiny\ensuremath{\bm{\leftrightarrow}}}}}{\mathbf{R}},\label{eq:polarizability-transformation}
\end{equation}
where

\begin{widetext}

\begin{equation}
\overset{\text{\text{\tiny\ensuremath{\bm{\leftrightarrow}}}}}{\mathbf{R}}\left(\theta,\phi,\chi\right)=\left(\begin{array}{ccc}
\mathrm{c}(\theta)\mathrm{c}(\phi)\mathrm{c}(\chi)-\mathrm{s}(\phi)\mathrm{s}(\chi) & \mathrm{c}(\theta)\mathrm{c}(\chi)\mathrm{s}(\phi)+\mathrm{c}(\phi)\mathrm{s}(\chi) & -\mathrm{c}(\chi)\mathrm{s}(\theta)\\
-\mathrm{c}(\chi)\sin(\phi)-\mathrm{c}(\theta)\mathrm{c}(\phi)\mathrm{s}(\chi) & \mathrm{c}(\phi)\mathrm{c}(\chi)-\mathrm{c}(\theta)\mathrm{s}(\phi)\mathrm{s}(\chi) & \mathrm{s}(\theta)\mathrm{s}(\chi)\\
\mathrm{c}(\phi)\mathrm{s}(\theta) & \mathrm{s}(\theta)\mathrm{s}(\phi) & \mathrm{c}(\theta)
\end{array}\right)
\end{equation}

\noindent is an orthogonal rotation matrix parametrized by Euler angles
as defined in \citep{Zare}. In the rotating reference frame, it is
convenient to choose a basis including the three principal axes of
inertia. In this basis, the polarizabilty tensor has a simple representation
\begin{equation}
\overset{\text{\text{\tiny\ensuremath{\bm{\leftrightarrow}}}}}{\boldsymbol{\alpha}}_{\mathrm{mol}}=\left(\begin{array}{ccc}
\alpha_{\perp} & 0 & 0\\
0 & \alpha_{\perp} & 0\\
0 & 0 & \alpha_{\parallel}
\end{array}\right).
\end{equation}

The explicit expression for $\braket{\overset{\text{\text{\tiny\ensuremath{\bm{\leftrightarrow}}}}}{\boldsymbol{\alpha}}}_{\mathrm{lab}}$
in terms of the Euler angles reads
\begin{equation}
\braket{\overset{\text{\text{\tiny\ensuremath{\bm{\leftrightarrow}}}}}{\boldsymbol{\alpha}}}_{\mathrm{lab}}=\left(\begin{array}{ccc}
A & D & 0\\
D & B & 0\\
0 & 0 & C
\end{array}\right),\label{eq:polarizability-lab-frame}
\end{equation}
where
\begin{align}
A & =\frac{1}{4}\left[\alpha_{\parallel}+3\alpha_{\perp}-\Delta\alpha\braket{\cos(2\theta)-2\cos(2\phi)\sin^{2}(\theta)}\right],\label{eq:App-A}\\
B & =\frac{1}{4}\left[\alpha_{\parallel}+3\alpha_{\perp}-\Delta\alpha\braket{\cos(2\theta)+2\cos(2\phi)\sin^{2}(\theta)}\right],\label{eq:App-B}\\
C & =\frac{1}{2}\left[\alpha_{\parallel}+\alpha_{\perp}+\Delta\alpha\braket{\cos(2\theta)}\right],\label{eq:App-C}\\
D & =\frac{\Delta\alpha}{2}\braket{\sin(2\phi)\sin^{2}(\theta)},\label{eq:App-D}
\end{align}

\noindent with $\Delta\alpha=\alpha_{\parallel}-\alpha_{\perp}$.
\end{widetext}

Pump pulse(s) initiate rotational dynamics, such that, generally,
the various expectation values appearing in Eqs. \eqref{eq:App-A},
\eqref{eq:App-B}, \eqref{eq:App-C}, and \eqref{eq:App-D} depend
on the probe delay. The evaluation of the elements of $\braket{\overset{\text{\text{\tiny\ensuremath{\bm{\leftrightarrow}}}}}{\boldsymbol{\alpha}}}_{\mathrm{lab}}$
requires simulating the rotational dynamics of laser driven linear
molecules. For the quantum mechanical simulations, we expressed the
trigonometric functions involved in the matrix elements of $\braket{\overset{\text{\text{\tiny\ensuremath{\bm{\leftrightarrow}}}}}{\boldsymbol{\alpha}}}_{\mathrm{lab}}$
in terms of Wigner D-functions as follows\begin{widetext}
\begin{align}
\cos(2\theta)-2\cos(2\phi)\sin^{2}(\theta) & =\frac{4D_{00}^{2*}-1}{3}-\sqrt{\frac{8}{3}}\left[D_{20}^{2*}+D_{-20}^{2*}\right],\\
\cos(2\theta)+2\cos(2\phi)\sin^{2}(\theta) & =\frac{4D_{00}^{2*}-1}{3}+\sqrt{\frac{8}{3}}\left[D_{20}^{2*}+D_{-20}^{2*}\right],\\
\cos(2\theta) & =\frac{4D_{00}^{2*}-1}{3},\\
\frac{1}{2}\sin(2\phi)\sin^{2}(\theta) & =i\sqrt{\frac{1}{6}}\left[D_{-20}^{2*}-D_{20}^{2*}\right].
\end{align}

\end{widetext}

\end{document}